\documentclass[12pt]{spieman}  
\usepackage{amsmath,amsfonts,amssymb}
\usepackage{graphicx}
\usepackage{setspace}
\usepackage{tocloft}
\usepackage{float}
\usepackage{lineno}

\title{Uniformity and Stability of the LSST Focal Plane}

\author[a]{Paul O'Connor}
\affil[a]{Brookhaven National Laboratory, Instrumentation Division, Upton, NY 11973}

\cftpagenumbersoff{figure}
\cftpagenumbersoff{table} 

\begin{document}
\maketitle

\begin{abstract}
    The LSST focal plane consists of 21 autonomous modules ("Raft Tower Modules", RTMs), each of which contains nine thick, fully-depleted 4K X 4K CCDs with associated control and readout electronics. To enable LSST's repetitive short-exposure cadence while maintaining high duty factor and low read noise, the readout is highly parallelized into 3024 independent video channels (16 per CCD, 144 per RTM). Two vendors supplied the LSST sensors; the devices have compatible mechanical and electrical interfaces and meet the same electro-optic specifications, but each RTM is constructed with sensors from a single supplier.
    The full complement of rafts were assembled at Brookhaven National Laboratory during January 2017 - January 2019. Each unit underwent extensive electro-optic and metrology characterization at operating temperature, the results of which are presented here along with a discussion of uniformity and stability.

\end{abstract}


\section {Introduction}

The LSST is a next-generation imaging instrument  with 320m$^2$-deg$^2$ etendue, designed to carry out a 10-year survey targeting dark matter and dark energy, solar system and Milky Way populations, and optical transients \cite{ivezic2008lsst}. The 3.2 Gpixel LSST camera, developed by a US Department of Energy collaboration, will have a science array of 189 fully-depleted CCDs arranged in 21 submodules called Raft Tower Modules (RTMs)\cite{o2012development}\cite{o2016integrated}, making it the largest digital camera thus far built for astronomical research. LSST's survey cadence involves covering the sky in repeated short exposures, making it necessary to minimize the closed-shutter readout time in addition to maximizing throughput across the 350-1050nm wavelength band. Table \ref {table:1} compares the LSST focal plane with other recent large imaging cameras.

\begin{table}[h]
\centering
\begin{tabular}{|l|l|l|l|l|l|}
\hline
Instrument & Camera & \begin{tabular}[c]{@{}l@{}}Science\\ CCDs\end{tabular} & \begin{tabular}[c]{@{}l@{}}Pixel\\ count\end{tabular} & \begin{tabular}[c]{@{}l@{}}Video\\ channels\end{tabular} & \begin{tabular}[c]{@{}l@{}}Readout\\ time\end{tabular} \\ \hline
Pan-STARRS 1 & GPC-1\cite{onaka2008pan} & 60 & 1.44G & 480 & 7s \\ \hline
Dark Energy Survey & DECam\cite{flaugher2015dark} & 62 & 504M & 124 & 20s \\ \hline
Subaru & HyperSuprime Cam\cite{miyazaki2012hyper} & 104 & 872M & 208 & 20s \\ \hline
LSST & LSSTCam\cite{kahn2010design} & 189 & 3.024G & 3024 & 2s \\ \hline
\end{tabular}
\caption{Imaging focal plane array comparison}
\label{table:1}
\end{table}

\begin{figure}[h]
    \centering
    \includegraphics[width=\textwidth]{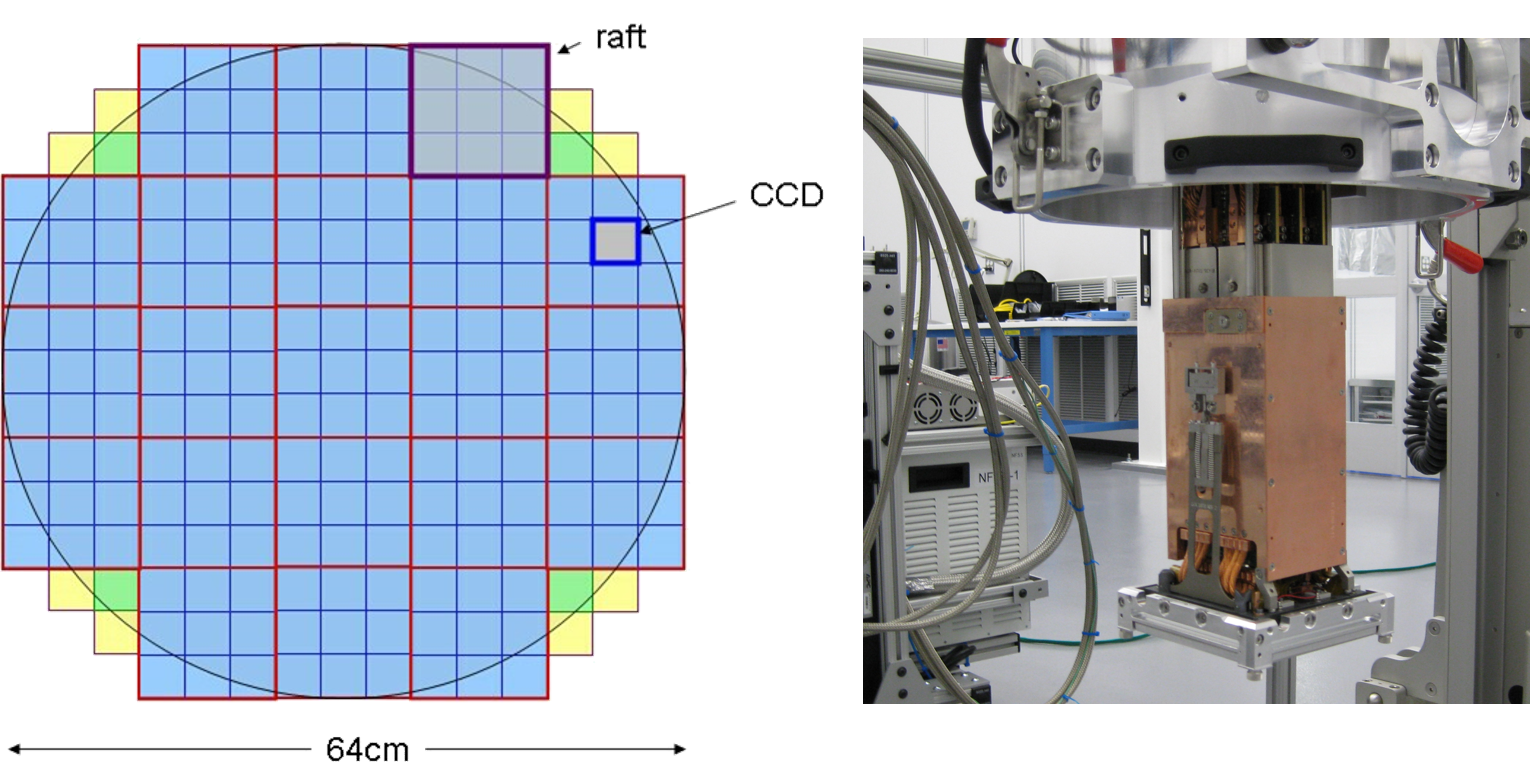}
    \caption{Left, arrangement of CCDs and RTMs in the LSST science focal plane.Right, assembled RTM being inserted into test cryostat. CCD subarray, downward-facing in this view, is covered by aluminum protective frame.}
    \label{fig:FPA+RTM}
\end{figure}

The science array is arranged as a set of 21 autonomous, fully testable modules each containing nine fully-depleted 4K X 4K CCDs together with all CCD control and video processing electronics contained in a compact, cryostat-compatible enclosure. A diagram of the focal plane layout and a photograph of one assembled RTM are shown in Figure \ref{fig:FPA+RTM}.

\subsection{Sensors and Electronics}
Key requirements for the LSST sensors were formulated early in the project, and a multi-year prototyping program was carried out with several suppliers. The production devices are of two types, the CCD-250 made by Teledyne-e2v (henceforth E2V) and the STA3800C, designed by Semiconductor Technology Associates, wafers fabricated at Teledyne-DALSA, and devices  postprocessed, packaged, and tested at the Imaging Technology Laboratory (ITL) of the University of Arizona. Both devices share a common 4K X 4K pixel format, 10 $\mu$m square pixels, with 16 independently-read out amplifier segments of 512 X 2K pixels. Devices are back-illuminated and fabricated on high-resistivity p-type silicon thinned to 100 $\mu$m for an optimal tradeoff of near-IR quantum efficiency and charge diffusion\cite{o2006study}\cite{radeka2009lsst}. Devices have 4-side buttable packages and achieve \textgreater90\% fill factor including non-imaging silicon area and inter-chip gaps. The two device types are 100\% interchangeable in their mechanical and electrical interfaces to the RTM, but differ in the following ways:

\begin{table}[h!]
\begin{tabular}{lll}
 & E2V & ITL \\
Parallel clock phases & 4 & 3 \\
Output amplifier & 2-stage & 1-stage \\
Entrance window &  implant & chemisorption  \\
Package style & cantilevered Si wirebonded & In bump bonded \\
Antiblooming stop & Yes & No \\
Tip/tilt/piston control & shims & glue-up with gauge blocks
\end{tabular}
\end{table}

Both devices are treated with proprietary coatings on the entrance side and the substrate-facing side, leading to differences in quantum efficiency at various wavelengths. Finally, ITL device outputs are buffered by JFET source followers mounted on the flex cables that interface the sensor to the RTM electronics boards. These were found to be necessary to provide sufficiently fast video rise and fall times to meet the 2s frame readout requirement.

To accommodate the high number of video channels in the LSST focal plane, each RTM incorporates a compact, ASIC-based control and readout electronics system\cite{russo2014lsst} on three PC boards occupying the $\sim$4-liter volume in the shadow of the CCD subarray. The RTM electronics includes 144 channels of video processing (amplification, dual-slope integration filtering, 18-bit digitization, data multiplexing and serial output link), CCD bias, timing, and control signal generation, thermal control of the CCD array, power conditioning, and monitoring and readback of several hundred temperatures, voltages, and currents. A strict power budget of \textless 50W (average) is necessary to match the heat removal capacity of the cryostat refrigeration system.

\subsection{Electro-optic Performance Requirements and Production Test Methods}
The performance requirements for the RTM are summarized in Table \ref{table: reqts+meas} below:

\begin{table}[h]
\centering
\begin{tabular}{|l|l|l|l|}
\hline
Parameter & \begin{tabular}[c]{@{}l@{}}Requirement\\ (threshold)\end{tabular} & \begin{tabular}[c]{@{}l@{}}Measured\\ (median of 21 RTMs)\end{tabular} & unit \\ \hline
QE u & $\geq$41 & 68 & \%  \\ \hline
QE g & $\geq$78 & 89.4 & \% \\ \hline
QE r & $\geq$88 & 95.2 &\%  \\ \hline
QE i & $\geq$81 & 98.4 & \% \\ \hline
QE z & $\geq$75 & 87.5 & \% \\ \hline
QE y & $\geq$21 & 29.9 & \% \\ \hline
Diffusion & $\leq$5.0 & 4.20 & um rms  \\ \hline
\begin{tabular}[c]{@{}l@{}}Dark current\\ (95th-percentile)\end{tabular} & $\leq$0.2 & 0.017 & e-/pix/s \\ \hline
Unusable pixels & $\leq$1.0 & 0.095 & \% \\ \hline
\begin{tabular}[c]{@{}l@{}}Frame read time\\ (144 Mpix)\end{tabular} & $\leq$2 & 1.94 & sec \\ \hline
Read noise & $\leq$9 (13) & 4.84 & e- rms \\ \hline
\begin{tabular}[c]{@{}l@{}}CTI serial\\ (at 1ke- signal)\end{tabular} & $\leq$5 (30) & 1.6 & ppm \\ \hline
\begin{tabular}[c]{@{}l@{}}CTI parallel\\ (at 1ke- signal)\end{tabular} & $\leq$3 & 0.6 & ppm\\ \hline
Power dissipation & 58.2 & 39 & W \\ \hline
Electronic crosstalk & 2 & 0.08 \emph{(typ.)} & \% \\ \hline
\end{tabular}
\caption{RTM electro-optic performance requirements and measured median performance for the 189 CCDs, 3024 video channels on 21 rafts. Numbers in parentheses are minimum requirements to meet the project science goals.}
\label{table: reqts+meas}
\end{table}

Since each RTM is able to function as standalone camera, the RTM electro-optic test stand has been built to simulate, as closely as possible, the conditions that will be experienced in the final LSST focal plane: early versions of the camera control\cite{marshall2006lsst} and data acquisition\cite{perazzo2007camera} software are used to execute exposure sequences, substantial use is made of the LSST data management software stack\cite{juric2015lsst} for image analysis, and prototype LSST power supplies are used. RTMs are housed in a vacuum cryostat with cold plates held at $-130^\circ$C and $-60^\circ$C to remove heat from the sensor array and electronics, respectively. (The LSST camera will hold the electronics cold plate at $-40^\circ$C, but our commercial closed-cycle cryocooler does not have sufficient cooling capacity at the higher temperature.)

The EO test methods follow the guidelines for CCD tests in [\citenum{doherty2013testmethods}]. CCD clock and bias voltages and timing sequences were set to the manufacturer's suggested values. In early raft testing some variations around the standard settings were explored; additional work to optimize performance (e.g., [\citenum{snyder2018optimization}]) is underway but is a separate activity from the production tests reported here.  Key measurements that are made are (1) \textsuperscript{55}Fe exposures, providing measurements of of gain (by fitting the Mn-K\textsubscript{$\alpha$} and K\textsubscript{$\beta$} lines for reconstructed clusters), noise (from overscan pixels), and charge diffusion (from cluster sizes); (2) dark exposures of 500s; (3) monochromatic flatfields from 350 to 1100nm for quantum efficiency; (4) monochromatic flatfield pairs at increasing exposure times for linearity, full well capacity, and photon transfer curve; (5) superflat exposures, coadded to produce low-noise files suitable for charge transfer efficiency measurements using the Extended Pixel Edge Response method\cite{janesick2001scientific}; and (6) twelve- to twenty-hour runs with continuous \textsuperscript{55}Fe exposures, to estimate response stability. Note that in [\citenum{doherty2013testmethods}], full well is defined as the maximum output signal level, not the signal level at which the charge transfer breaks down. Also, in keeping with general practice the "gain" of a channel is expressed in e-/ADU (actually an inverse gain). For most rafts, EO runs were carried out at two CCD temperatures ($-90^\circ$C and $-100^\circ$C); results given in subsequent sections are for $-90^\circ$C temperature.

The tests described here have been developed to verify, using fully automated acquisition and analysis, that the LSST science rafts satisfy the performance criteria in Table \ref{table: reqts+meas}. Further studies of subtle characteristics (persistence, crosstalk, distortions due to static and dynamic electrostatic effects, etc.) have been carried out on single CCDs and are reported in the references\cite{astier2015introduction}\cite{o2015crosstalk}, \cite{antilogus2014brighter}\cite{guyonnet2015evidence}.

\section{Uniformity Results}
\subsection{Quantum Efficiency (QE)}
\begin{figure}[h]
    \centering
      \includegraphics[width=\textwidth, height=10cm]{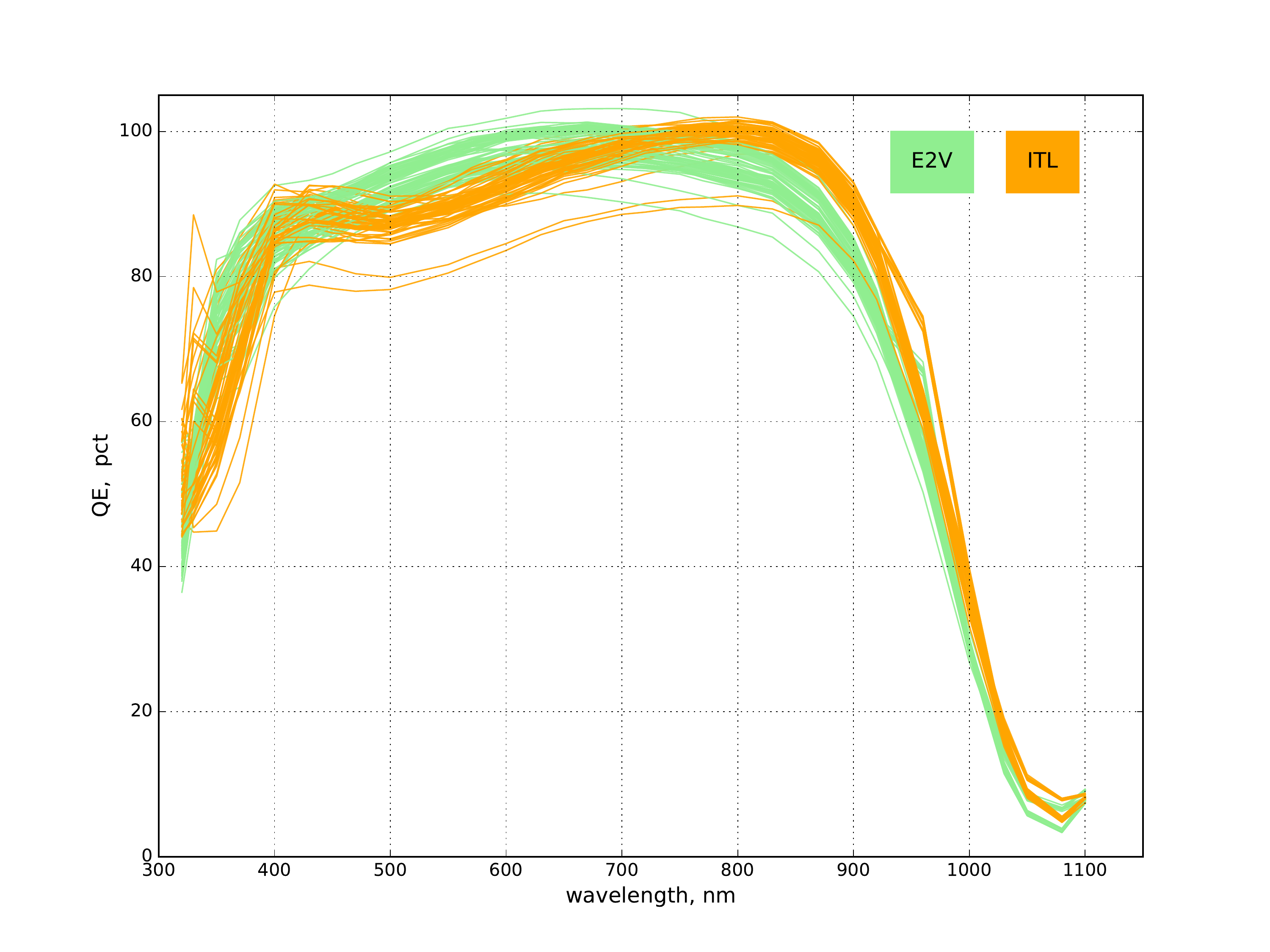}
    \caption{QE vs wavelength for E2V (green) and ITL (orange) sensors (72 sensors of each type).}
    \label{fig:QEcurves}
\end{figure}

QE measurement uses conventional methods\cite{coles2017automated} and is referenced to a NIST-calibrated photodiode and corrected for flatfield irradiance falloff across the raft surface. A single number is reported for each CCD by averaging the response after correcting the individual segments' gain and offset. Photoresponse nonuniformity is typically at the 1\% level in midband, up to 3\% rms below 450nm and above 950nm due to window processing and fringing respectively. Figure \ref{fig:QEcurves} shows the QE curves for all CCDs, separated by supplier type. Absolute QE numbers have uncertainties of 2--5\%, while relative variations include instrumental drifts over the 28-month production period. In general, there is higher QE at low and high wavelengths for E2V and ITL sensors, respectively.

\subsection{Noise, Full Well, Dark Current, Charge Diffusion, Charge Transfer Inefficiency}
\begin{figure}[h]
    \centering
    \includegraphics[width=\textwidth]{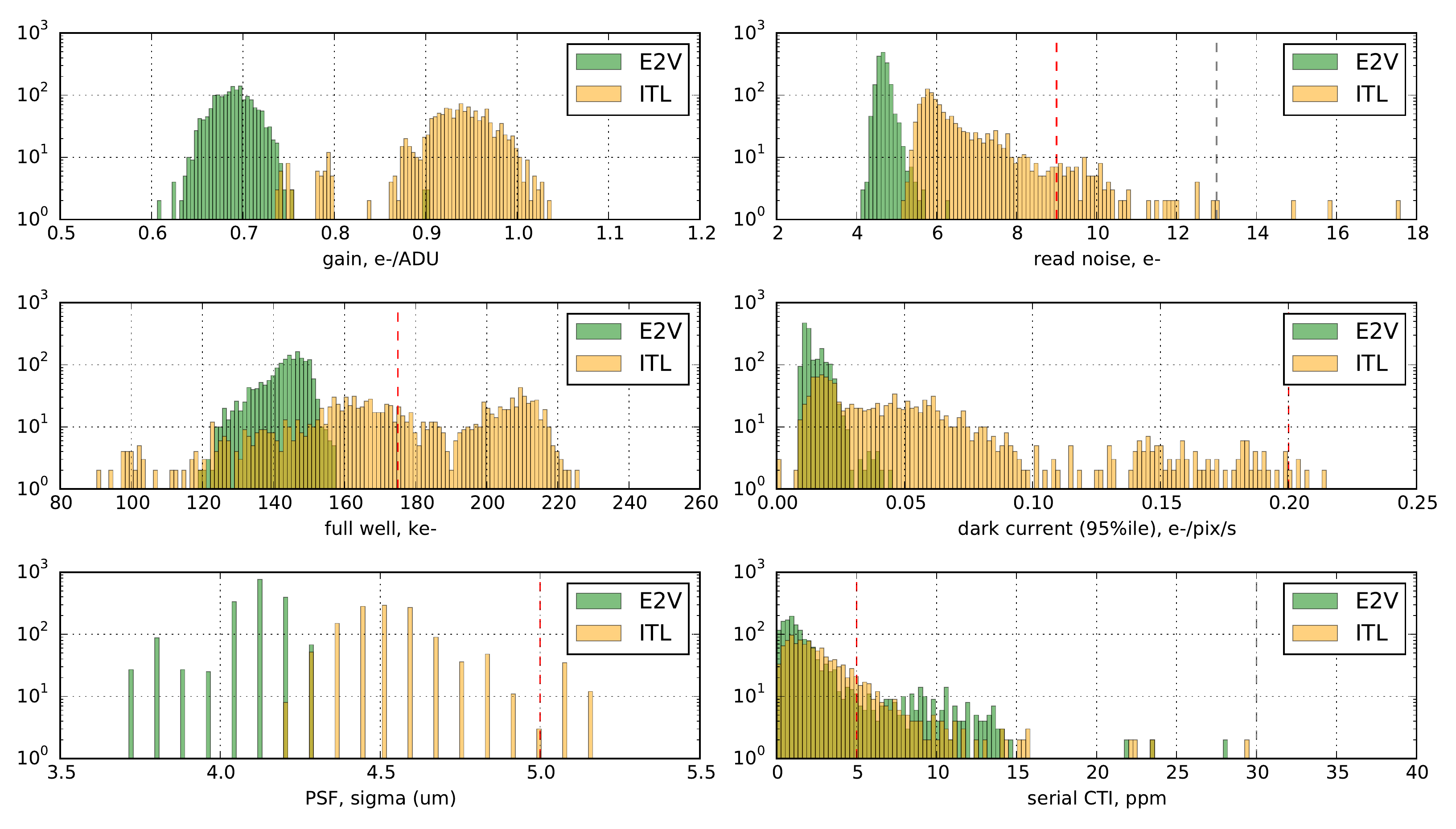}
    \caption{Histograms of electro-optic parameters; E2V (green) and ITL (orange). Dashed lines indicate hard (red) and soft (grey) specification limits from Table \ref{table: reqts+meas}. Note: Charge diffusion PSF measurement uses coarsely-quantized bins.}
    \label{fig:EOhist}
\end{figure}

Figure \ref{fig:EOhist} shows the distribution of parameters for the 189 CCDs (3024 channels) of the focal plane, separated by CCD supplier. Population statistics are summarized in Table \ref{table:EOPopStats}.

\begin{figure}[h]
    \centering
    \includegraphics[width=\textwidth]{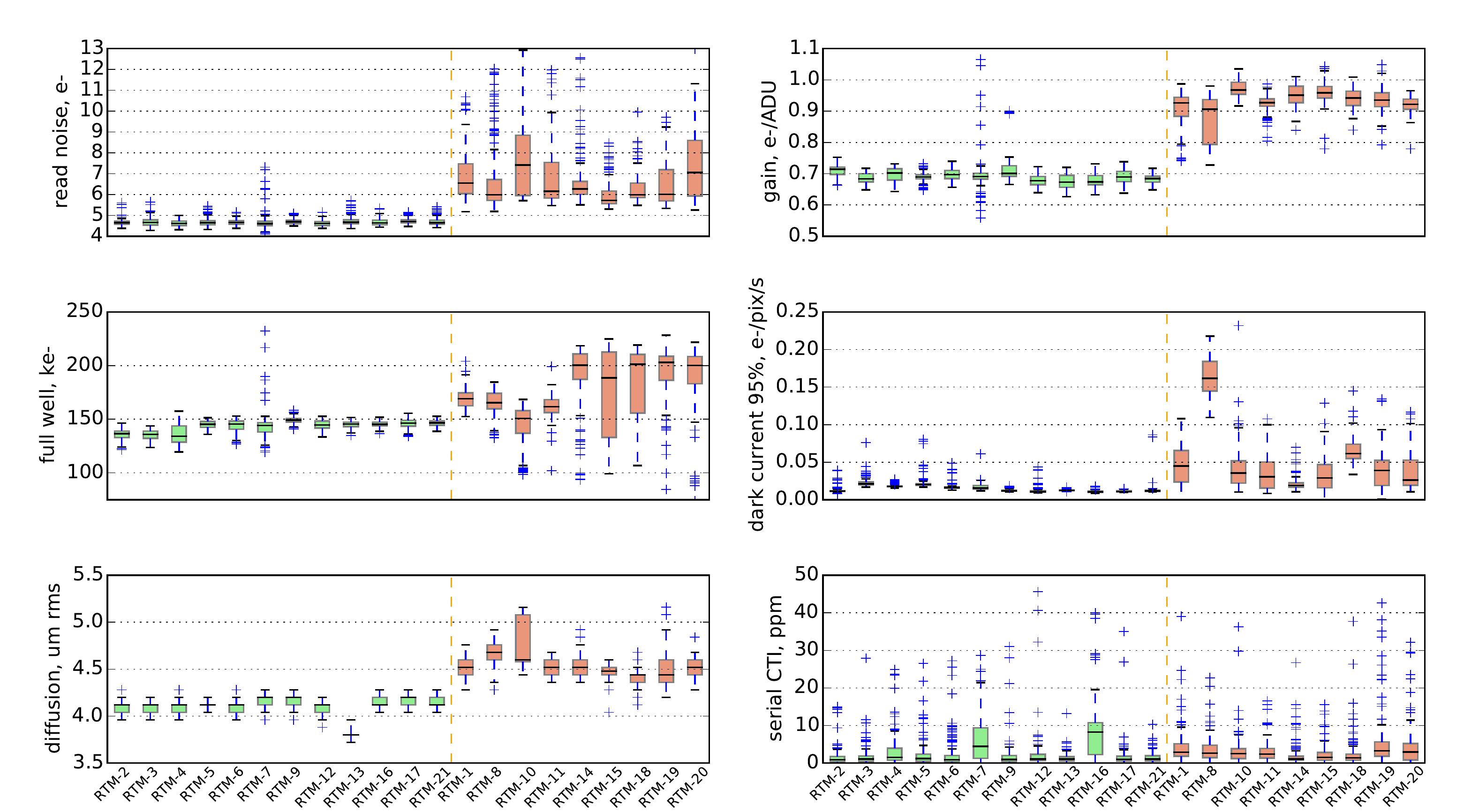}
    \caption{RTM-by-RTM electro-optic parameters for E2V (green) and ITL (orange) rafts.}
    \label{fig:EObox}
\end{figure}

\begin{figure}[h]
    \centering
    \includegraphics[width=\textwidth, height=6cm]{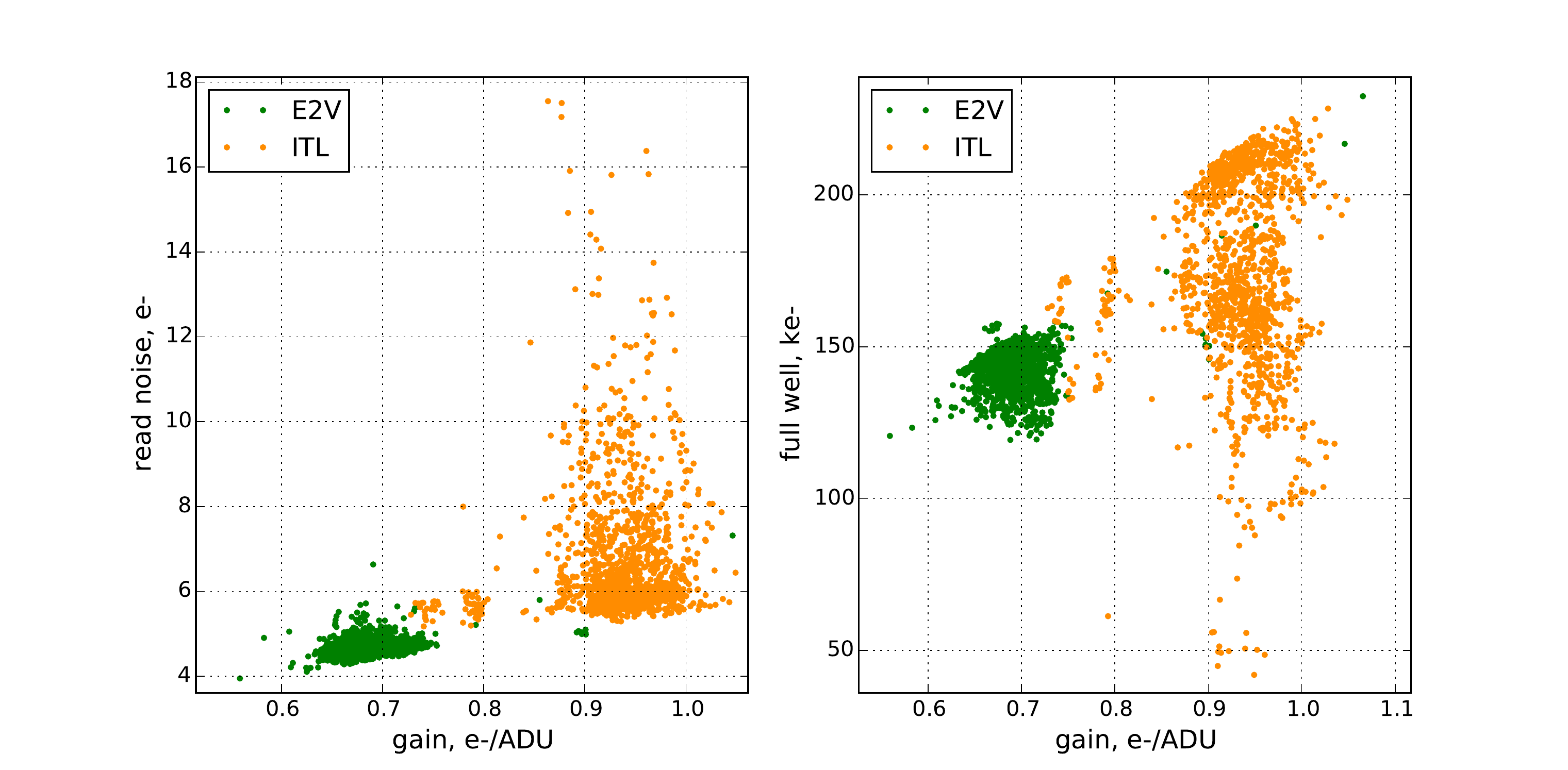}
    \caption{Correlation between gain and read noise (left), full well capacity (right).}
    \label{fig:gainCorr}
\end{figure}

In contrast to the QE results, the electro-optic parameters show a greater difference between CCD types, with nearly non-overlapping distributions in some cases. Furthermore, the ITL sensors show greater variability in parameters. Figure \ref{fig:gainCorr} shows that the read noise and full well parameter differences are linked through the different gain of the two CCD types. Gain, read noise, and full well are all  determined by the sense node capacitance\cite{janesick2001scientific} which apparently differs by roughly 35\% between the two suppliers. The shape of the full well vs. gain plot suggests that at least for some channels, the electronics dynamic range rather than actual CCD properties may be limiting. There are straightforward ways to adjust the electronics gain during operation to ensure that the maximum CCD signal is always measured, as this is necessary for accurate crosstalk correction.

\begin{table}[ht]
\centering
\begin{tabular}{|l|l|l|l|l|l|l|l|}
\hline
Parameter, $P$ & \(\overline{E2V}\) & \(\overline{ITL}\) & \(\sigma_{E2V}\) & \(\sigma_{ITL}\) & \(dP/PdT_{E2V}\) & \(dP/PdT_{ITL}\) & unit (tempco) \\ \hline

QE-u & 69.8 & 62.2 & 6.3 & 7.8 & 0.13 & 0.50 & \% (\%/$^\circ$C) \\ \hline
QE-g & 89.9 & 88.4 & 2 & 3.4 & 0.027 & -0.023 & \% (\%/$^\circ$C) \\ \hline
QE-r & 95.6 & 94.3 & 2.4 & 3.4 & -0.017 & -0.011 & \% (\%/$^\circ$C) \\ \hline
QE-i & 95.1 & 99.4 & 3.6 & 4.4 & -0.016 & -0.016 & \% (\%/$^\circ$C) \\ \hline
QE-z & 84.2 & 92.7 & 2.4 & 4.6 & 0.088 & 0.063 & \% (\%/$^\circ$C) \\ \hline
QE-y & 25.9 & 31.5 & 3.5 & 4.2 & 0.51 & 0.80 & \% (\%/$^\circ$C) \\ \hline
read noise & 4.7 & 6.1 & 0.23 & 1.8 & -0.035 & 0.09 & e- rms (\%/$^\circ$C) \\ \hline
gain & 0.69 & 0.94 & 0.034 & 0.11 & 0.023 & 0.020 & e-/ADU (\%/$^\circ$C) \\ \hline
full well & 144 & 186 & 8 & 40 & - & - & ke- (\%/$^\circ$C) \\ \hline
dark current 95\% & 0.013 & 0.038 & 0.03 & 0.048 & 0.52 & -0.20 & e-/pix/s (\%/$^\circ$C) \\ \hline
Diffusion PSF & 4.12 & 4.48 & 0.14 & 0.28 & - & - & um rms (\%/$^\circ$C) \\ \hline
CTI-serial & 1.52 & 2.07 & 6.6 & 45 & - & - & ppm (\%/$^\circ$C) \\ \hline
\end{tabular}
\caption{Population statistics and temperature coefficients for all LSST Science Rafts.}
\label{table:EOPopStats}
\end{table}

\subsection{Correlated noise, and effect of reduced readout rate}
\begin{figure}[h]
    \centering
    \includegraphics[width=\textwidth]{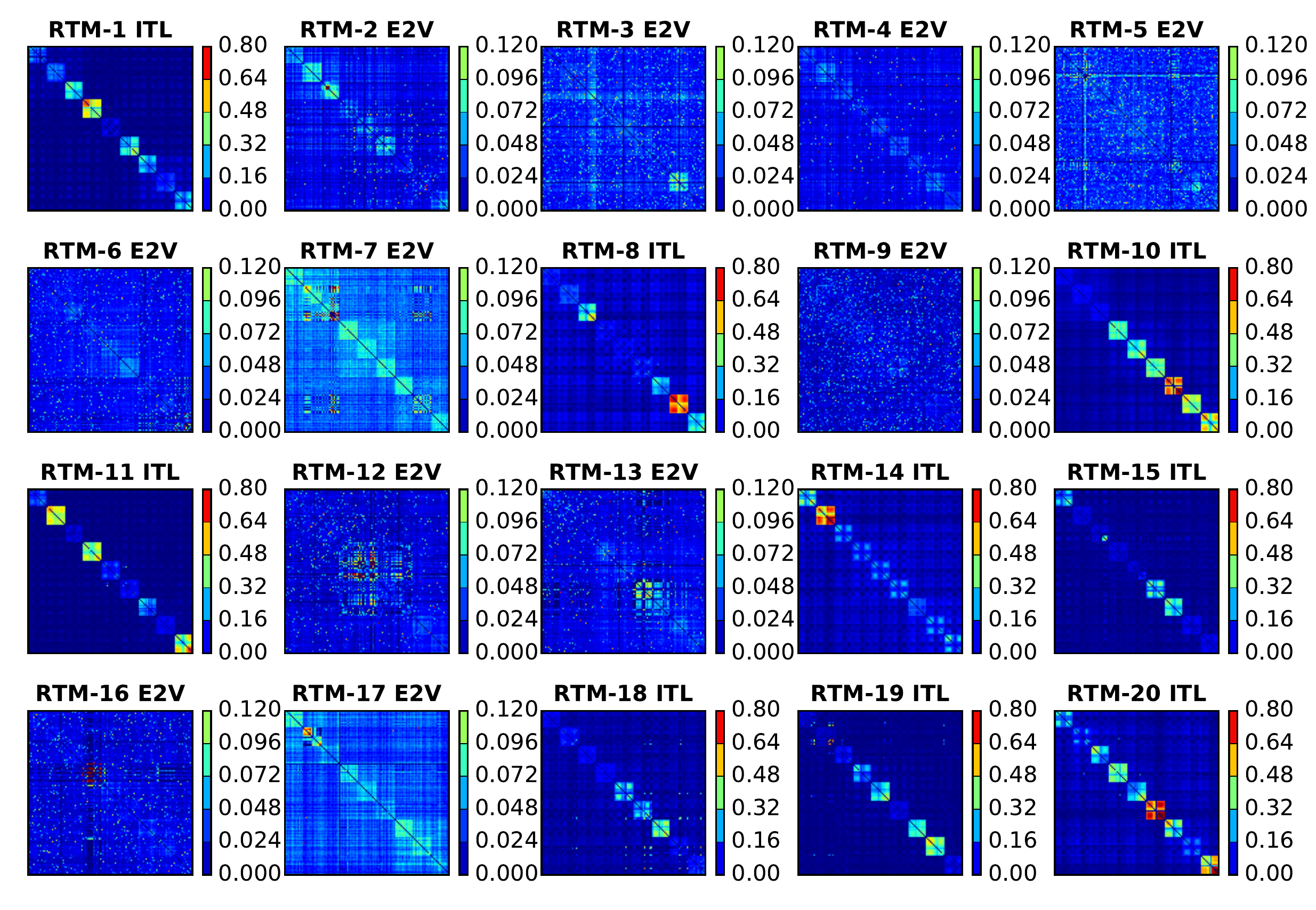}
    \caption{Noise correlation matrices for 20 RTMs, at 2s frame readout time. In these plots, the correlation coefficient between all 144x144 channel pairs is shown. The 16x16 block diagonals show the intra-CCD coefficients; the larger 48x48 block diagonals represent the correlations within a single electronics board. Note different scales for E2V and ITL rafts}
    \label{fig:corCoef}
\end{figure}

\begin{figure}[h]
    \centering
    \includegraphics[width=\textwidth]{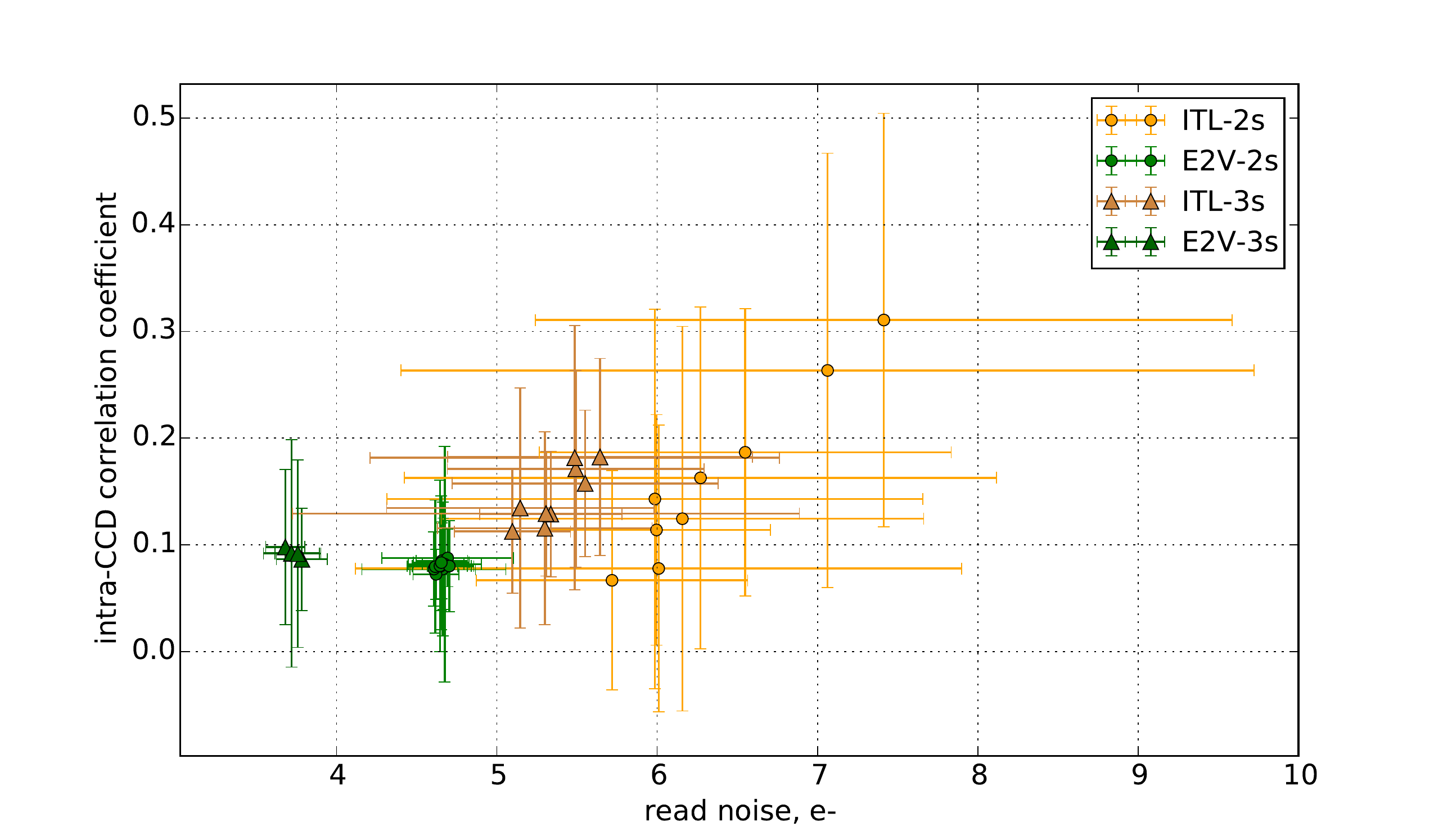}
    \caption{Read noise (x-axis) and intra-CCD correlation for 13 RTMs at 2s and 3s frame readout times.}
    \label{fig:RN-CC_relation}
\end{figure}

At the pixel rate required to read out the CCD in 2 seconds, clock transitions necessarily occur close to CDS integration intervals. We made two observations which suggested that clock-coupled noise was present in some devices. First, clock feedthrough was seen to be pronounced in video waveforms of the noisier devices. Second, noise correlation measurements were made (see Figure \ref{fig:corCoef}) which clearly showed individual devices with significant correlation between the noise waveforms of the 16 channels in bias images. Comparison of the degree of intra-CCD correlation and read noise (Figure \ref{fig:RN-CC_relation}) confirmed the connection. 

\begin{figure}[h]
    \centering
    \includegraphics[width=\textwidth]{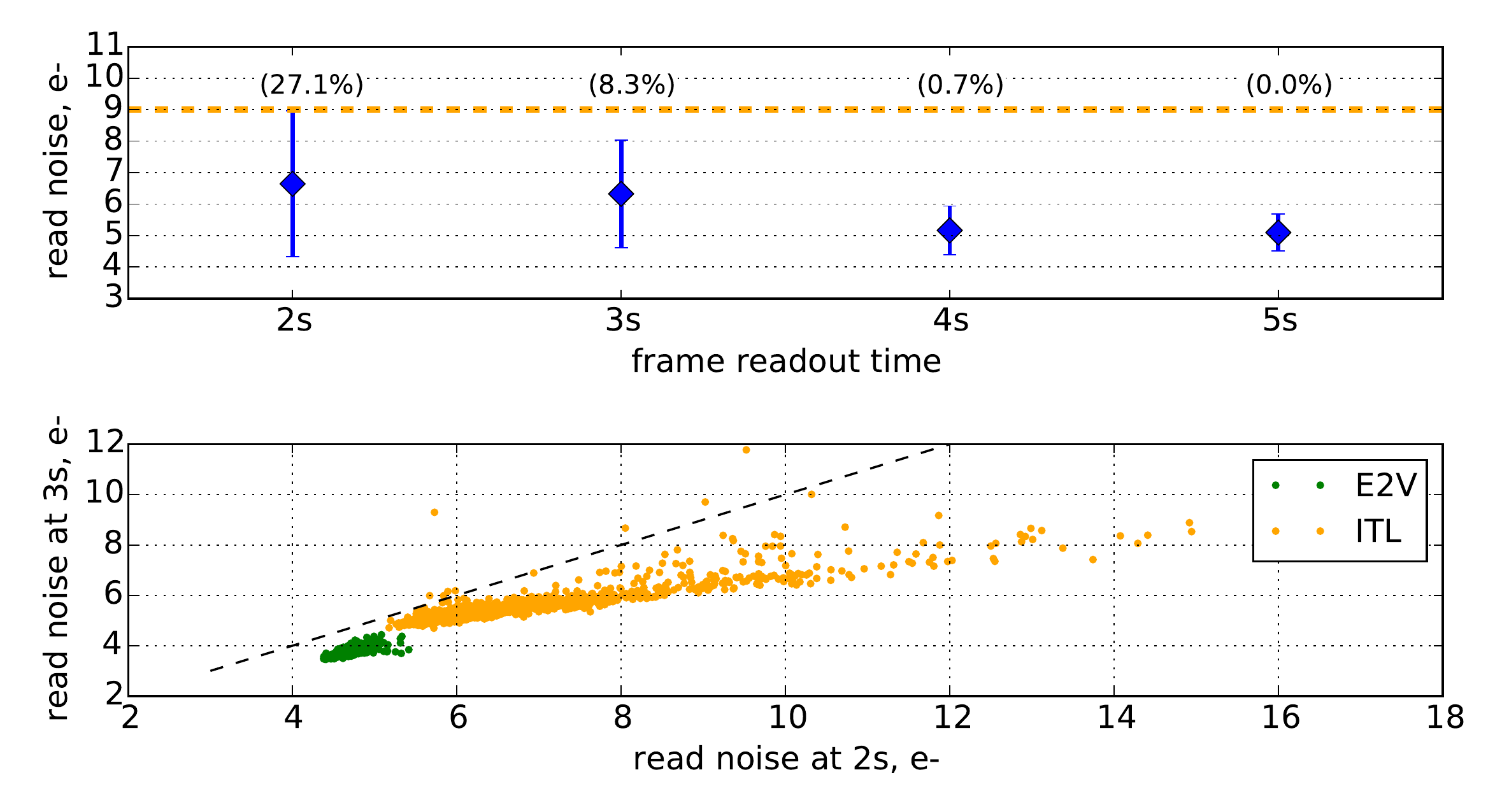}
    \caption{Top: RTM-10 read noise vs. frame readout time. Numbers in parentheses are percent of channels with noise exceeding 9e-. Bottom: Noise results for 13 rafts at 2s and 3s readout time.}
    \label{fig:Noise_vs_Frametime}
\end{figure}

We observed that read noise could be reduced by adding delays between the clock edges and the integration intervals, while keeping the integration times constant. For RTM-10, we varied the timing to give frame readout times of 2, 3, 4, and 5 seconds; both the noise and the dispersion in noise decreased (Figure \ref{fig:Noise_vs_Frametime}, top). Thirteen rafts were then measured at both 2s and 3s readout times. An improvement of 10-20\% was found for the slower readout, with the noisiest channels on ITL devices benefiting the most (Figure \ref{fig:Noise_vs_Frametime}, lower).

\section{Stability Results in Test Cryostat}

\begin{figure}[h]
    \centering
    \includegraphics[width=\textwidth]{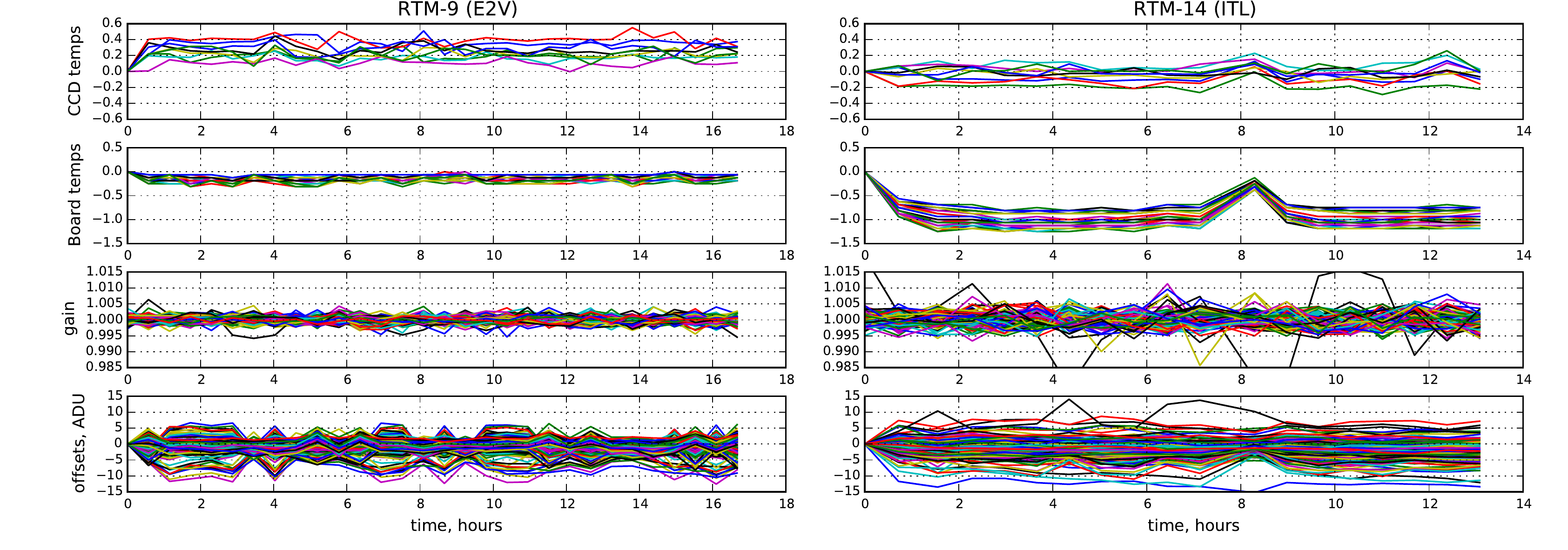}
    \caption{Timeseries of CCD tmeperatures, electronics temperatures, gains, and offsets for RTM-6 (E2V) and RTM-14 (ITL). 144-channel gains are normalized to their mean value throughout the run, temperatures and offsets are shown as differences from their initial values.}
    \label{fig:GShistory2Rafts}
\end{figure}

\begin{figure}[h]
    \centering
    \includegraphics[width=\textwidth]{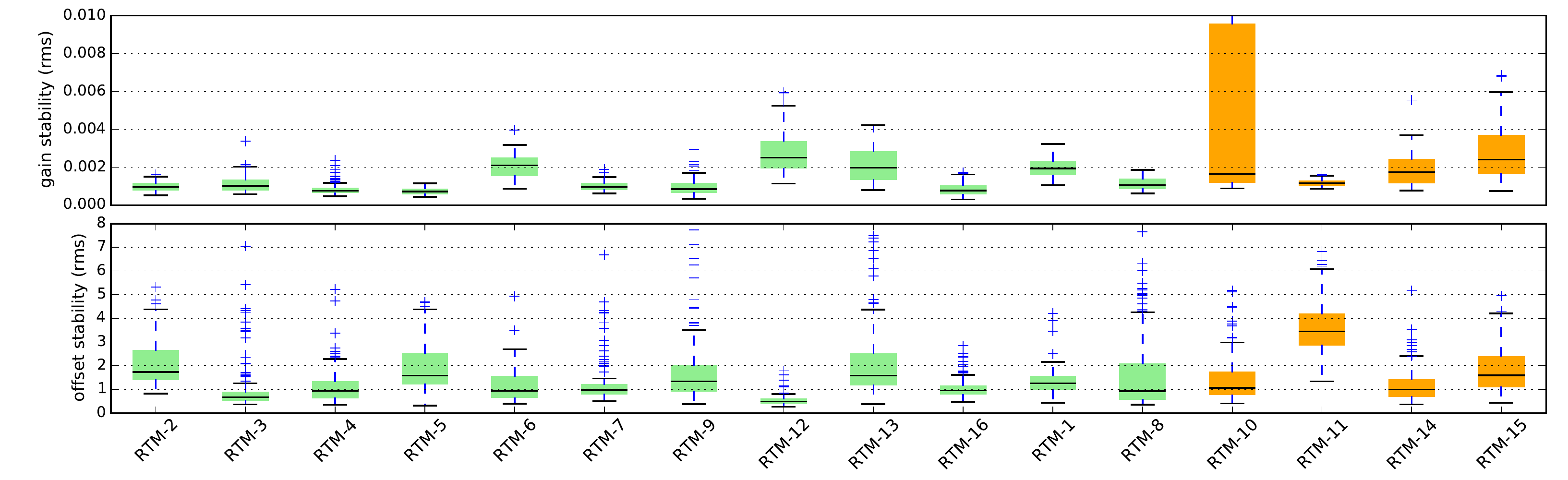}
    \caption{Long-term stability results for gain (upper) and offset (lower) of all channels for 12 E2V (green) and 6 ITL (orange) rafts.}
    \label{fig:GSresults}
\end{figure}

For each raft, a long-duration run of \textsuperscript{55}Fe exposures was executed. CCD temperatures were controlled by the RTM's internal thermal control loop, while the electronics were cooled by conduction to a cold plate stabilized to -60$^\circ$C. On average, temperatures in the RTM were stable to $0.15 \pm 0.08$ and   $0.22 \pm 0.22$ $^\circ$C for the CCDs and raft electronics, respectively. Representative time series histories for two RTMs are given in Figure \ref{fig:GShistory2Rafts}. In Figure \ref{fig:GSresults} we show the gain and offset stability range for the 144 channels of each of 16 rafts. After taking statistical error in the gain determination and EMI noise in the temperature readout into account, the gain variation seen in Fig. \ref{fig:GShistory2Rafts} is consistent with the temperature coefficients in Table \ref{table:EOPopStats}.

\section{Discussion}
\begin{figure}[h]
    \centering
    \includegraphics[height=8cm]{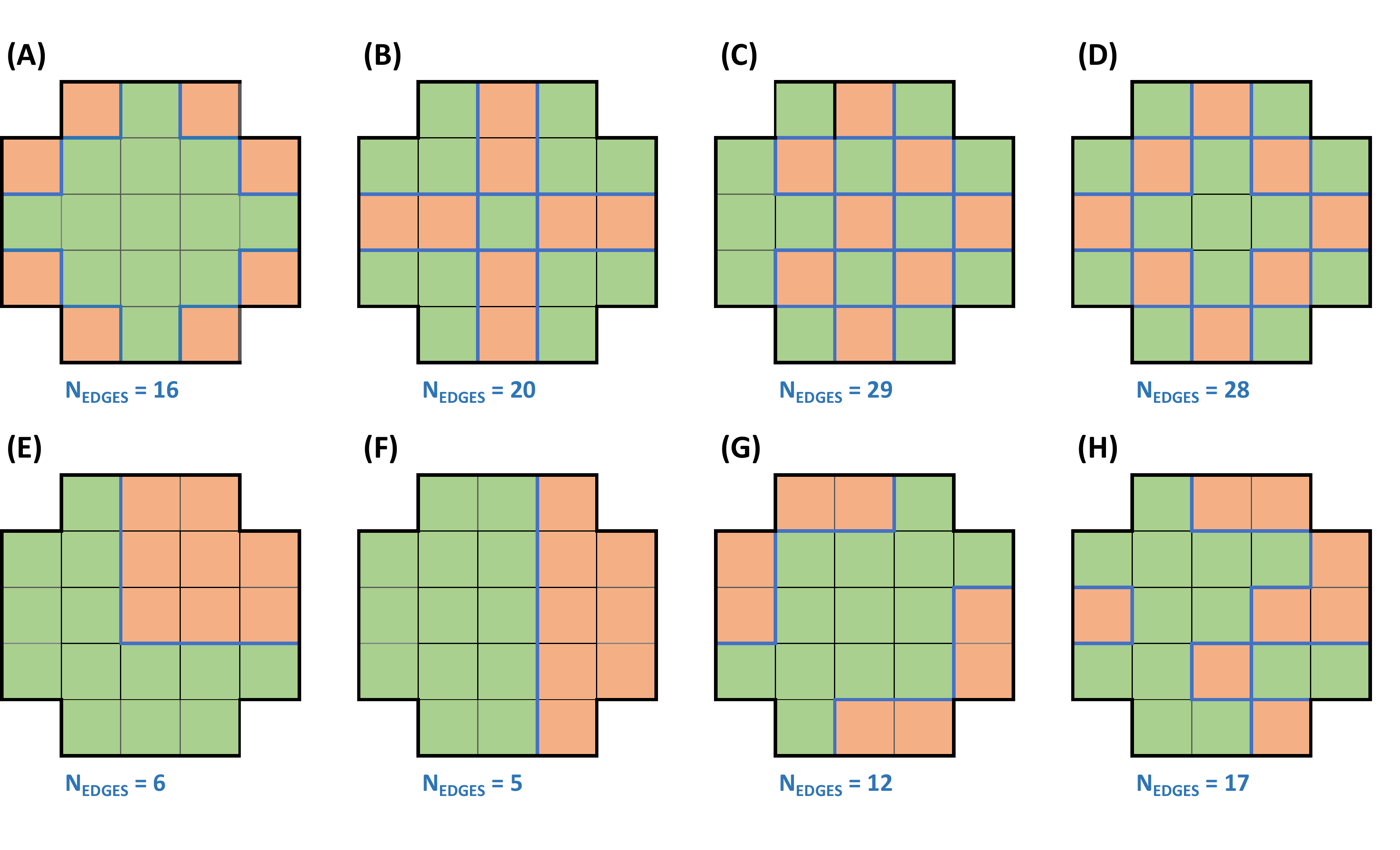}
    \caption{Alternative configurations of populating the LSST FPA with 13 E2v (green) and 8 ITL (orange) rafts; number of dissimilar-sensor edges shown in blue.}
    \label{fig:SplitFPA}
\end{figure}

\begin{figure}[h]
    \centering
    \includegraphics[width=\textwidth]{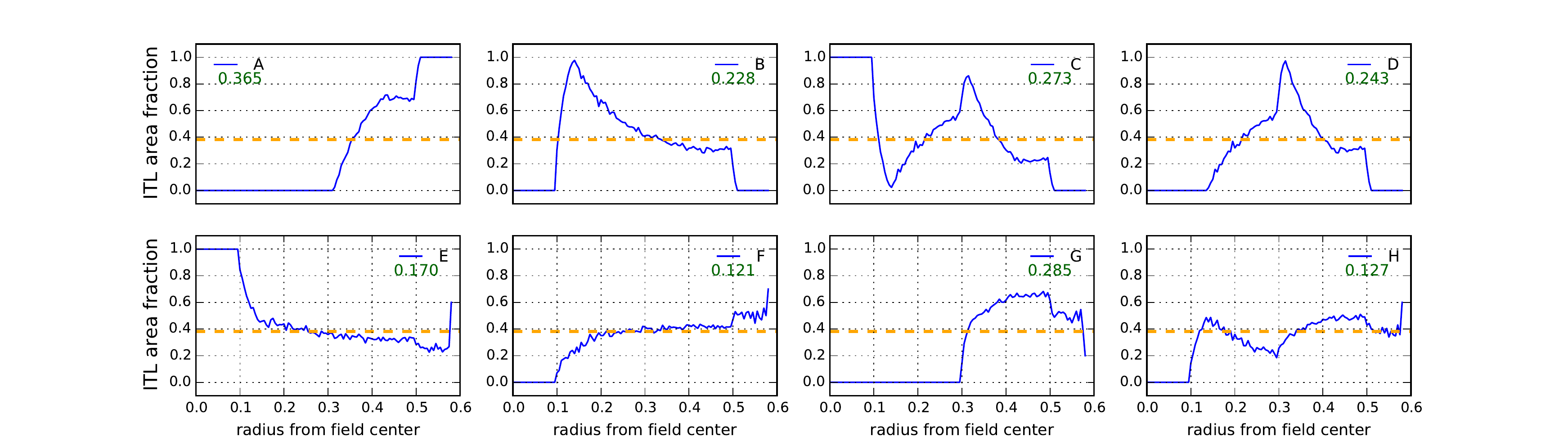}
    \caption{Azimuthally-averaged area fraction occupied by ITL rafts, for the configurations of Fig. \ref{fig:SplitFPA}. Normalized departure from complete uniformity shown in green for each configuration.}
    \label{fig:ITLFrac}
\end{figure}

References [\citenum{PST2015MixedFPA}] and [\citenum{Slater2017}]   discuss science impacts of a mixed-sensor focal plane. The science impacts considered in [\citenum{PST2015MixedFPA}] include photo-z measurement systematics and artificial structure imprinted on the galaxy power spectrum by the spatially-nonuniform FPA, while [\citenum{Slater2017}] discussed the impact on transient detection. Both references consider only the wide-fast-deep (WFD) survey where the impact of mixed-FPA nonuniformity is mitigated by the dithering of field centers and sky rotation angles. However, in the Deep Driling (DD) fields, repeated exposures will likely be acquired with random sky rotations but only small translational dithering of the field centers. This will result in radial variations in the probability that a DD source will be seen by either sensor type, with the degree of variation depending on the distribution of rafts by supplier type across the focal plane. In Fig. \ref{fig:SplitFPA} we show eight possible arrangements of placing the 8 ITL and 13 E2V rafts in the focal plane. In Fig. \ref{fig:ITLFrac} we show the azimuthally-averaged probability that an object at radius \emph{r} will be observed by an ITL sensor in the DD fields (blue lines), compared with the FPA-averaged probability of 8/13 (dashed horizontal line). 

Another issue not considered in [\citenum{PST2015MixedFPA}] is PSF interpolation across dissimilar sensor boundaries. It is likely that the accuracy of PSF interpolation will be higher when PSF calibrators are located on the same sensor type as shape measurement targets. Figure \ref{fig:SplitFPA} shows the number of dissimilar-sensor edges for the eight selected configurations. From Figures \ref{fig:SplitFPA} and \ref{fig:ITLFrac} it can be seen that configurations E and F  minimize both the radial uniformity variation and the number of  dissimilar-sensor boundaries compared to the other arrangements.

\section{Conclusions}
All 21 focal plane modules (rafts) for the 3 Gpixel LSST science focal plane have been constructed and evaluated for electro-optic performance. Population statistics for the main CCD parameters (including temperature coefficients) have been measured. For the ensemble of rafts, median EO performance meets requirements with margin. Dispersion of the EO parameters is strongly tied to the CCD supplier, with the largest differences seen in gain, read noise, full well capacity, and charge diffusion. Increasing the readout time from 2 to 3s improves the read noise considerably, and minimizes the difference between the two sensor types. Gain and offset stability were measured by acquiring x-ray images over 12+ hours. Gains remained stable at the 0.1 - 0.2\% level while offset variation of only 1-2 electrons was observed. The number of dissimilar-CCD boundaries and the azimuthally-averaged distribution of the two sensor types about the field center is determined by the placement of the 13 + 8 rafts in the LSST cryostat.

\clearpage

\section{ACKNOWLEDGEMENTS}
Homer Neal, James Chiang, and Claire Juramy-Giles contributed code for EO data acquisition and analysis. The BNL Science Raft team led by William Wahl maintained and operated the test stand. Their contributions are gratefully acknowledged.

This manuscript has been co-authored by employees of Brookhaven Science Associates, LLC.,
Portions of this work are supported by the Department of Energy under contract DE-SC0012704
with Brookhaven National Laboratory. LSST project activities are supported in part by the National
Science Foundation through Governing Cooperative Agreement 0809409 managed by the
Association of Universities for Research in Astronomy (AURA), and the Department of Energy
under contract DE-AC02-76-SFO0515 with the SLAC National Accelerator Laboratory. Additional
LSST funding comes from private donations, grants to universities, and in-kind support from
LSSTC Institutional Members.

\bibliography{JATIS_20190131}   

\begin{thebibliography}{10}

\bibitem{ivezic2008lsst}
Z.~Ivezic, J.~Tyson, B.~Abel, {\em et~al.}, ``Lsst: from science drivers to
  reference design and anticipated data products,'' {\em arXiv preprint
  arXiv:0805.2366}   (2008).

\bibitem{o2012development}
P.~O'Connor, I.~Kotov, P.~Takacs, {\em et~al.}, ``Development of the lsst raft
  tower modules,'' in {\em High Energy, Optical, and Infrared Detectors for
  Astronomy V},   {\bf 8453}, 84530L, International Society for Optics and
  Photonics  (2012).

\bibitem{o2016integrated}
P.~O'Connor, P.~Antilogus, P.~Doherty, {\em et~al.}, ``Integrated system tests
  of the lsst raft tower modules,'' in {\em High Energy, Optical, and Infrared
  Detectors for Astronomy VII},   {\bf 9915}, 99150X, International Society for
  Optics and Photonics  (2016).

\bibitem{onaka2008pan}
P.~Onaka, J.~Tonry, S.~Isani, {\em et~al.}, ``The pan-starrs gigapixel camera\#
  1 and stargrasp controller results and performance,'' in {\em Ground-based
  and Airborne Instrumentation for Astronomy II},   {\bf 7014}, 70140D,
  International Society for Optics and Photonics  (2008).

\bibitem{flaugher2015dark}
B.~Flaugher, H.~Diehl, K.~Honscheid, {\em et~al.}, ``The dark energy camera,''
  {\em The Astronomical Journal} {\bf 150}(5), 150  (2015).

\bibitem{miyazaki2012hyper}
S.~Miyazaki, Y.~Komiyama, H.~Nakaya, {\em et~al.}, ``Hyper suprime-cam,'' in
  {\em Ground-based and Airborne Instrumentation for Astronomy IV},   {\bf
  8446}, 84460Z, International Society for Optics and Photonics  (2012).

\bibitem{kahn2010design}
S.~Kahn, N.~Kurita, K.~Gilmore, {\em et~al.}, ``Design and development of the
  3.2 gigapixel camera for the large synoptic survey telescope,'' in {\em
  Ground-based and Airborne Instrumentation for Astronomy III},   {\bf 7735},
  77350J, International Society for Optics and Photonics  (2010).

\bibitem{o2006study}
P.~O'Connor, V.~Radeka, D.~Figer, {\em et~al.}, ``Study of silicon thickness
  optimization for lsst,'' in {\em High Energy, Optical, and Infrared Detectors
  for Astronomy II},   {\bf 6276}, 62761W, International Society for Optics and
  Photonics  (2006).

\bibitem{radeka2009lsst}
V.~Radeka, J.~Frank, J.~Geary, {\em et~al.}, ``Lsst sensor requirements and
  characterization of the prototype lsst ccds,'' {\em Journal of
  Instrumentation} {\bf 4}(03), P03002  (2009).

\bibitem{russo2014lsst}
S.~Russo, P.~Antilogis, H.~Lebbolo, {\em et~al.}, ``The lsst science raft
  electronics,'' in {\em RT2014-19th Real-Time Conference},   (2014).

\bibitem{marshall2006lsst}
S.~Marshall, J.~Thaler, T.~Schalk, {\em et~al.}, ``Lsst camera control
  system,'' in {\em Advanced Software and Control for Astronomy},   {\bf 6274},
  627422, International Society for Optics and Photonics  (2006).

\bibitem{perazzo2007camera}
A.~Perazzo, R.~Herbst, M.~Huffer, {\em et~al.}, ``Camera data acquisition for
  the large synoptic survey telescope,'' in {\em Real-Time Conference, 2007
  15th IEEE-NPSS},  1--2, IEEE  (2007).

\bibitem{juric2015lsst}
M.~Juri{\'c}, J.~Kantor, K.~Lim, {\em et~al.}, ``The lsst data management
  system,'' {\em arXiv preprint arXiv:1512.07914} {\bf 1}  (2015).

\bibitem{doherty2013testmethods}
P.~Doherty and S.~Digel, ``Ccd sensor electro-optical testing methods,''
  internal technical memo, LSST Project  (2013).

\bibitem{snyder2018optimization}
A.~Snyder, K.~Gilmore, and A.~Roodman, ``Optimization of ccd operating voltages
  for the lsst camera,'' in {\em High Energy, Optical, and Infrared Detectors
  for Astronomy VIII},   {\bf 10709}, 107092B, International Society for Optics
  and Photonics  (2018).

\bibitem{janesick2001scientific}
J.~R. Janesick {\em et~al.}, {\em Scientific charge-coupled devices}, vol.~117,
  SPIE press Bellingham  (2001).

\bibitem{astier2015introduction}
P.~Astier, ``An introduction to some imperfections of ccd sensors,'' {\em
  Journal of Instrumentation} {\bf 10}(05), C05013  (2015).

\bibitem{o2015crosstalk}
P.~O'Connor, ``Crosstalk in multi-output ccds for lsst,'' {\em Journal of
  Instrumentation} {\bf 10}(05), C05010  (2015).

\bibitem{antilogus2014brighter}
P.~Antilogus, P.~Astier, P.~Doherty, {\em et~al.}, ``The brighter-fatter effect
  and pixel correlations in ccd sensors,'' {\em Journal of Instrumentation}
  {\bf 9}(03), C03048  (2014).

\bibitem{guyonnet2015evidence}
A.~Guyonnet, P.~Astier, P.~Antilogus, {\em et~al.}, ``Evidence for
  self-interaction of charge distribution in charge-coupled devices,'' {\em
  Astronomy \& Astrophysics} {\bf 575}, A41  (2015).

\bibitem{coles2017automated}
R.~Coles, J.~Chiang, D.~Cinabro, {\em et~al.}, ``An automated system to measure
  the quantum efficiency of ccds for astronomy,'' {\em Journal of
  Instrumentation} {\bf 12}(04), C04014  (2017).

\bibitem{PST2015MixedFPA}
S.~Ritz and {LSST Project Science Team}, ``Camera mixed focal plane option,
  report-241,'' internal technical memo, LSST Project  (2015).

\bibitem{Slater2017}
C.~Slater, R.~Jones, E.~Bellm, {\em et~al.}, ``Impact of a heterogeneous focal
  plane on lsst image differencing, ldm-523,'' internal technical memo, LSST
  Project  (2017).

\end{thebibliography}
\bibliographystyle{spiejour}   

\end{document}